%% file: master_springer.tex
\documentclass[review]{elsarticle}
\usepackage[a4paper, left=4cm, right=4cm, top=4cm, bottom=4cm]{geometry}

\usepackage[english]{babel}
\usepackage[utf8]{inputenc}
\usepackage[T1]{fontenc}
\usepackage{color}
\usepackage{lmodern, microtype}

\usepackage{amsmath}
\usepackage{amstext}
\usepackage{amssymb}
\usepackage{latexsym}
\usepackage{dsfont}
\usepackage{mathrsfs}
\usepackage{setspace}
\usepackage{ulem}
\usepackage{bbold}

\usepackage{hyperref}
\hypersetup{
    colorlinks=true,
    unicode=true
}

\usepackage{graphicx}
\usepackage{float}
\usepackage{caption, booktabs}
\usepackage{graphics}
\usepackage{subcaption}
\usepackage{lipsum}
\usepackage{amsmath,amssymb,amsfonts,amsthm}
\usepackage[table, dvipsnames]{xcolor}
\usepackage{mathrsfs}
\usepackage{soul}

\usepackage{threeparttable}

\definecolor{forest}{HTML}{0003c2}

\definecolor{lila}{HTML}{6e00c2}

\definecolor{cyan}{rgb}{0,0.5,.5}
\definecolor{dcyan}{rgb}{0,0.3,.3}

\definecolor{leaf}{HTML}{009e1d}

\usepackage{tikz}
\usepackage{lineno}
\modulolinenumbers[0]

\begin{document}


\begin{frontmatter}

    \journal{Contributions to Statistics}

    \title{Forecasting Natural Gas Prices with
Spatio-Temporal Copula-based Time Series
Models}


    \author[1]{Sven Pappert\corref{cor1}}
    \ead{pappert@statistik.tu-dortmund.de}
    \cortext[cor1]{Corresponding author}

    \author[1,2]{Antonia Arsova}
    \ead{arsova@statistik.tu-dortmund.de}

    \address[1]{Chair of Econometrics, Department of Statistics, TU Dortmund University, Germany}
    \address[2]{RWI -- Leibniz Institute for Economic Research, Germany}

    \begin{abstract}
        Commodity price time series possess interesting features, such as heavy-tailedness, skewness, heteroskedasticity, and non-linear dependence structures. These features pose challenges for modeling and forecasting. In this work, we explore how spatio-temporal copula-based time series models can be effectively employed for these purposes. We focus on price series for fossil fuels and carbon emissions. Further, we illustrate how the t-copula may be used in conditional heteroskedasticity modeling. The possible emergence of non-elliptical probabilistic forecasts in this context is examined and visualized. The problem of finding an appropriate point forecast given a non-elliptical probabilistic forecast is discussed. We propose a solution where the forecast is augmented with an artificial neural network (ANN). The ANN predicts the best (in MSE sense) quantile to use as point forecast. In a forecasting study, we find that the copula-based models are competitive.
    \end{abstract}
    \begin{keyword}
        Commoditiy Prices \sep Copula-based time series \sep Conditional Volatility \sep Forecasting \sep Vine Copula
    \end{keyword}
\end{frontmatter}

\newpage
\section{Introduction}
Modeling and forecasting commodity prices is important for trading, political decision making and economic adjustments. Especially in recent times, forecasting natural gas prices has gained importance following the russian invasion of Ukraine. In this work we focus on modeling and forecsting short-term natural gas future prices jointly with related commodity prices. We model the time series jointly to exploit the additional information carried by their mutual dependence. To this aim we employ spatio-temporal copula based time series models.\\
Copulas are popular choices to model the cross-sectional dependence in time series with the copula-GARCH approach in financial markets \cite{hu2006dependence, jondeau2006copula} as well as in energy markets \cite{aloui2014dependence, berrisch2022modeling}. In the copula-GARCH approach the temporal dependence of each time series is modeled by typical time series models, such as ARMA-GARCH. The cross-sectional dependence structure of the time series can be captured by finding the copula of the standardized residuals of the univariate time series model. 
However, such models are only able to allow for flexible dependence structures in the cross-sectional dimension. The mean process is modeled linearly.\\
On the other hand, temporal copula modeling (or 'copula-based time series modeling') as well as spatio-temporal copula time series modeling offers an alternative to classical linear time series approaches. The models are able to flexibly model cross-sectional as well as temporal dependencies. There is emerging literature on the topic. Chen \& Fan \cite{chen2006estimation} investigate the estimation of copula-based semiparametric time series models. The authors provide conditions for $\beta$-mixing and prove consistency as well as asymptotic normality using the Delta method. Beare \cite{beare2010copulas} further investigates mixing conditions. Smith et al. \cite{smith2010modeling} decompose serial dependence of intraday electricity load using pair copula constructions. Simard and R{\'e}millard  \cite{simard2015forecasting} investigate the forecasting performance of the spatio-temporal t-copula dependent on the strength and structure of the dependence as well as the marginal distributions. Beare \& Seo \cite{beare2015vine} as well as Nagler et al. \cite{nagler2022stationary} examine spatio-temporal vine copula models.\\
Examples where flexible dependence modeling can be important are the following. The cross-sectional dependence between international stock markets can be asymmetric with dominant lower tail dependence, indicating the phenomenon of contagion in financial markets. This was investigated by Hu \cite{hu2006dependence}. It was found that the asymmetric dependence dominates. With regard to energy markets it was found by Aloui et al. \cite{aloui2014dependence} that crude oil and gas markets rather comove during bullish periods. Thereby also displaying an asymmetric cross-sectional dependence structure. A concise example for a possible emergence of non-linear dependence structures in the temporal domain and hence requiring sophisticated dependence modeling is given in the introduction of the work by Beare \cite{beare2010copulas}. The continous growth of financial time series contrasted with their sudden and quick decrease represent an asymmetric temporal relation. Thus cross-sectional as well as temporal dependence modeling can be important in many fields. \\
In this paper we explore the possibilities and performances of spatio-temporal copula models for modeling energy market time series. The basic idea underlying spatio-temporal time series modeling with copulas is a decomposition of the joint distribution. Using Sklars theorem, \cite{sklar1959fonctions} the joint distribution of consecutive observations is decomposed into dependence and marginal structure, $F_{\mathbf{X}_t, \mathbf{X}_{t-1}}(\mathbf{a}, \mathbf{b}) = C\left[F_{\mathbf{X}_t}(\mathbf{a}), F_{\mathbf{X}_{t-1}}(\mathbf{b})\right]$. Various copula specifications can be employed. In this work, the t-copula \cite{demarta2005t} and the gaussian copula are considered as basic copula models for spatio-temporal time series modeling.
Vine copula models \cite{aas2009pair, czado2010pair} are also considered. Spatio-temporal forecasting with the t-copula was examined in \cite{simard2015forecasting}. The spatio-temporal vine copula modeling of multivariate time series is, among others, explored in \cite{beare2015vine, nagler2022stationary, smith2010modeling}.\\
Using the notion of conditional copulas, the models can be used for forecasting.
The resulting probabilistic forecasts can be non-elliptical. It is not obvious what constitutes a sensible point forecast in this case. The expectation value is a sensible point forecast for elliptical or almost elliptical probabilistic forecasts. For non-elliptical forecasts, e.g. a bimodular probabilistic forecast, the expecation value predicts points that are unlikely. One possible solution to this problem would be to take the mode of the probabilistic forecast as the point forecast.
Another possibility is to augment the forecasting procedure by an artificial neural network (ANN)\footnote{We refer to \cite{higham2019deep} for a concise introduction.} The ANN predicts which quantile of the probabilistic forecast is optimal (in MSE sense) as point forecast. The inputs of the ANN are past values of the times series and the last optimal quantiles. One advantage of the ANN-augmented forecast is that the ANN can be estimated and used for prediction completely independent from the probabilistic time series model estimation and forecast. It is well known that ANNs are very powerful with regards to point forecasting. In this approach the ANN point forecasts are also equipped with an underlying probabilistic distribution, enabling the calculation of confidence intervals and other distributional properties.   
The models performances are examined in a forecasting study.
A model closely related to the model from \cite{berrisch2022distributional}, which was shown to outperform other popular models, is considered as benchmark. We find that the spatio-temporal copula time series modeling with ANN-augmented point forecasts are competitive for natural gas and related commoditiy prices forecasting.\\ 
The main contributions of this paper are two-fold. The first contribution is the application oriented exploration of spatio-temporal (vine) copula time series models for the energy market. We evaluate the performance of the models in a forecasting study and find that they perform well. The second contribution is the methological exploration of point forecasting from non-elliptical probabilistic forecasts and the inclusion of ANN-augmented forecasts.\\
In the next Section, Sect.~\ref{Data}, the data used in this work is introduced briefly. Sect.~\ref{Methods} describes the statistical methods used in this work. The empirical results of the forecasting study are presented in Sect.~\ref{Results}, while Sect.~\ref{Conclusion} summarizes the results and outlines some avenues for future research. 
\section{Data description}
\label{Data}
\begin{figure}[!b]
\centering
\includegraphics[scale=0.5]{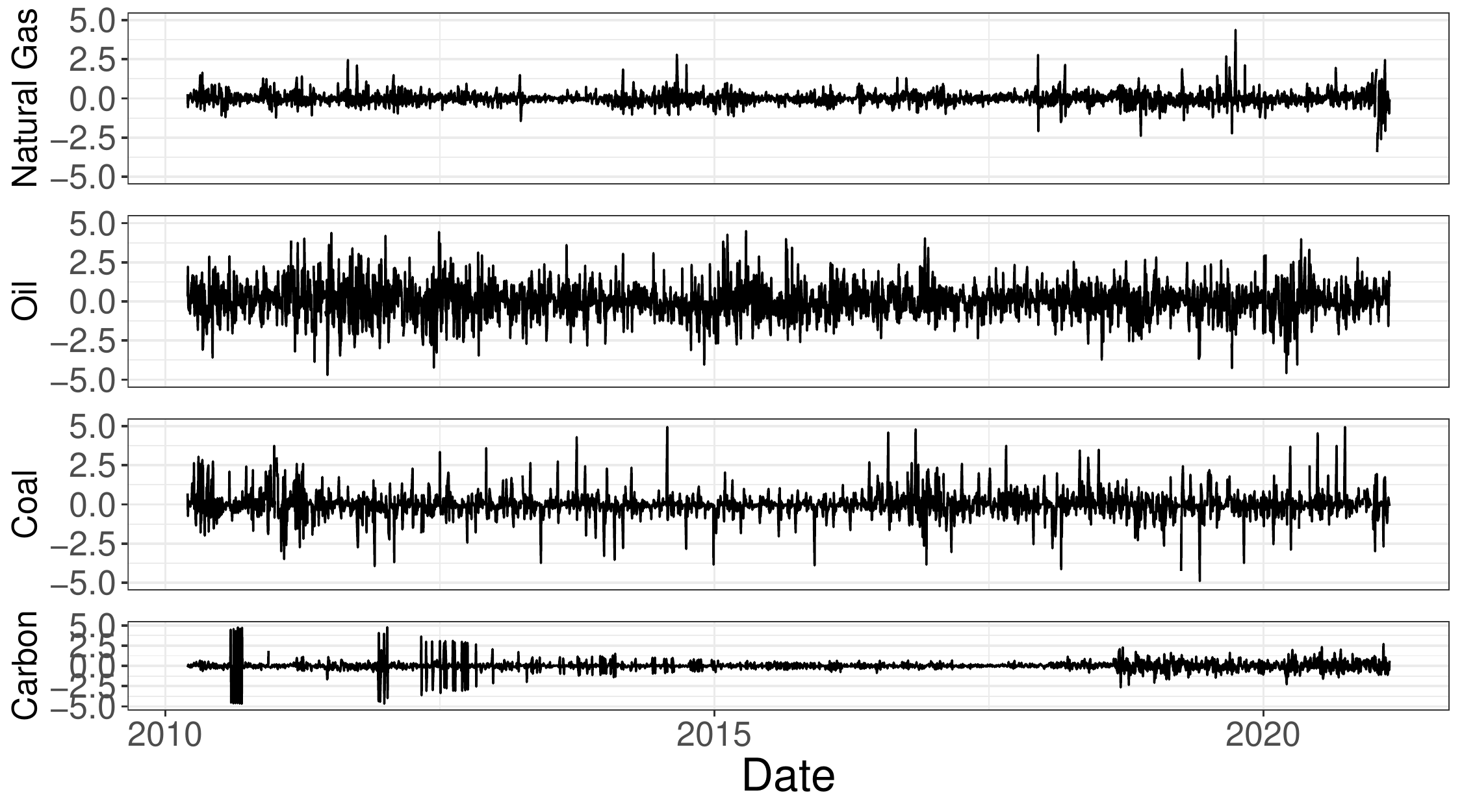}
\caption{First differences of the respective commodity price time series.}
\label{Data_first_diff}
\end{figure}

The time series analyzed in this work are extracted from the web-platform \url{investing.com} using the \texttt{Python} package \texttt{investpy} \cite{del2021investpy}. We analyze month-ahead natural gas futures (NGas) from the Netherlands (TTF Hub). The related commodities used for modeling are short-term carbon emission futures (CEF), short term brent oil futures (oil) and short term coal futures (coal). The analyzed time series are comprised of daily observations. The obervation period spans from March 2010 to February 2021. In total, the time series comprises 2861 observations.
Missing values, which occur especially during the holidays are trivially imputated as the last known value. The original time series are non-stationary. To obtain stationarity, which is necessary for the methods used in this report, the time series are differenced once. The differenced time series are displayed in Fig.~\ref{Data_first_diff}. The hypothesis of non-stationarity in the first differences is rejected by the Dickey-Fuller test at the level $\alpha=0.01$ for all time series.
\section{Statistical Methods}
\label{Methods}
This section comprises the description of the methods used in this report. First copulas and related notions are introduced. The copula specifications used in the analysis are presented and discussed.
The application of copulas to time series modeling follows. The emergence of non-elliptical probabilistic forecasts is examined with regard to the t-copula. It is shown how the t-copula may be used to model conditional heteroskedasticity. The need for new point forecast methods is presented. 
\subsection{Copulas}
\begin{figure}[!b]
\centering
\includegraphics[scale=0.5]{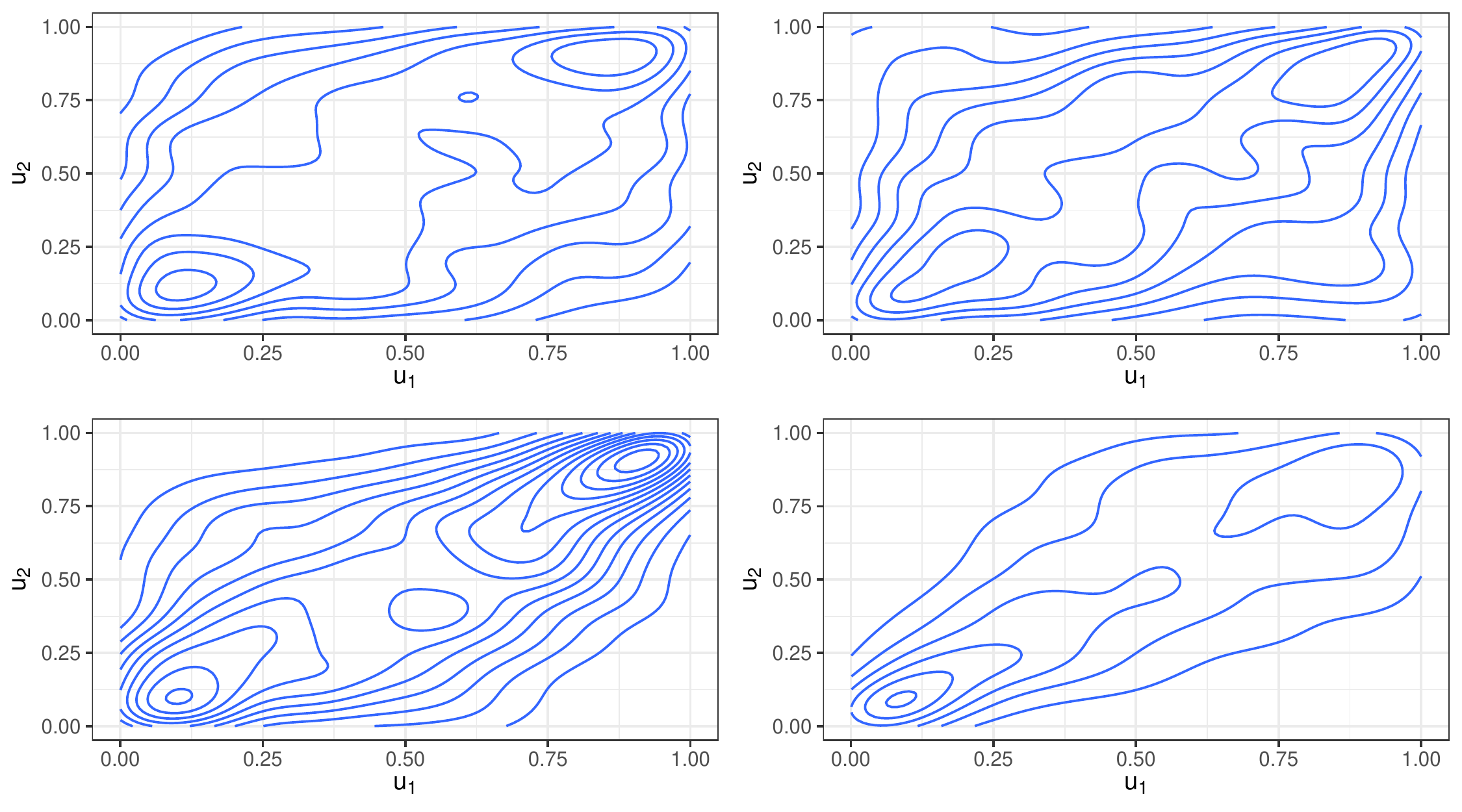}
\caption{Simulated density plots of four different two-dimensional copula specifications. The upper left plot shows the gaussian copula density with dependence parameter set to $\rho = 0.4$. Upper right shows the t-copula density with dependence parameter $\rho = 0.4$ and degree of freedom $\nu = 4$. The lower left plot displays the Gumbel copula density with dependence parameter $\theta = 2$. The lower right plot displays the Clayton copula density with dependence parameter $\theta = 2$. All plots were created with simulations with 2000 samples.}
\label{copula_simulations}
\end{figure}

Copulas are distribution functions on the unit cube with uniform marginals:
\begin{equation}
C: [0, 1]^d \rightarrow [0,1].
\end{equation}
Copulas gain their relevance by Sklars Theorem \cite{sklar1959fonctions}. It states that every multivariate distribution can be decomposed into a copula and marginal distributions. Let $X_{1}, \hdots, X_{d}$ be real valued random variables with joint distribution $F_{X_1, \hdots, X_d}$ and marginal distributions $F_{X_1}, \hdots, F_{X_d}$. Then it holds that there exists a copula $C$ such that
\begin{equation}
F_{X_1, \hdots, X_d}(x_1, \hdots, x_d) = C_{U_1, \hdots, U_d}\left[F_{X_1}(x_1), \hdots, F_{X_d}(x_d) \right],
\label{Sklar}
\end{equation}
where $(U_1, \hdots, U_d) := (F_{X_1}(X_1), \hdots, F_{X_d}(X_d))$. In the following the indices of the copula will be dropped. If the random variables $X_1, \hdots, X_d$ are continous then the decomposition is unique \cite{joe2014dependence}. The pseudo-observation $(u_1, \hdots, u_d) := F_{X_1}(x_1), \hdots, F_{X_d}(x_d)$ are, by virtue of the probability integral transformation, realizations from a uniform distribution, $U_i = F_{X_i}(X_i) \sim U[0,1], i \in \{1, \hdots, d \}$ \cite{angus1994probability}. This permits the copula to be interpreted as the dependence structure of the random variables $X_{1}, \hdots, X_{d}$. The copula density, $c$ which couples the joint density $f_{X_1, \hdots, X_d}$ and marginal densities $f_{X_1}, \hdots, f_{X_d}$ can be derived directly from Eq.~\ref{Sklar} by taking derivatives,
\begin{eqnarray}
f_{X_1, \hdots, X_d}(x_1, \hdots, x_d) =& c\left[F_{X_1}(x_1), \hdots, F_{X_d}(x_d)\right] f_{X_1}(x_1) \hdots f_{X_d}(x_d),
\\
c[u_1, \hdots, u_d] =& \frac{\partial^d C[u_1, \hdots, u_d]}{\partial u_1 \hdots \partial u_d}.
\end{eqnarray}
The copula density is important for estimation via maximum likelihood as well as for the visualization of dependence structures. In this paper the copula density will also be used to introduce the notion of vine copula models.
Another important notion for dependence modeling is the conditional copula of $U_1,\hdots,U_i$ given $U_{i+1}, \hdots, U_{d}$, respectively the conditional copula density. The conditional copula (density) can also be derived from Eq.~\ref{Sklar}. It is given by \cite{simard2015forecasting},
\begin{eqnarray}
C[u_1, \hdots, u_i | u_{i+1}, \hdots, u_d] =& \frac{\partial_{u_{i+1}} \hdots \partial_{u_d} C[u_1, \hdots, u_d]}{c[u_{i+1}, \hdots, u_{d}]},
\label{Conditional_Copula}
\\
c[u_1, \hdots, u_i | u_{i+1}, \hdots, u_d] =& \frac{c[u_1, \hdots, u_d]}{c[u_{i+1}, \hdots, u_{d}]}.
\label{Conditional_Copula_density}
\end{eqnarray}
Conditional copulas are especially relevant for conditional time series models as presented in this paper. The relation between the conditional density and the copula is as follows,
\begin{eqnarray}
f_{X_1, \hdots, X_i | X_{i+1} \hdots X_d}(x_1, \hdots, x_i | x_{i+1}, \hdots, x_d) =& \frac{c\left[F_{X_1}(x_1), \hdots, F_{X_d}(x_d)\right]}{c[u_{i+1}, \hdots, u_{d}]}
\\
& \times f_{X_1}(x_1) \hdots f_{X_i}(x_i). \nonumber
\label{conditional_density_copula}
\end{eqnarray}
The copula approach to multivariate modeling allows for separate modeling of marginal properties and dependence structure. This feature renders the approach far more flexible than standard multivaritate modeling. Joint distributions such as the multivariate normal or students t-distribution restrict the choice of marginal distributions. In the copula approach, marginal distributions can be arbitrary. Also the dependence structure of random variables can have various features, that have to be accounted for by choosing an appropriate copula specification. In this work, the gaussian, Clayton, Gumbel and t-copula are utilized. In the following they will be introduced briefly as the joint distribution of random variables $U_i \sim U[0,1], i \in \{1,\hdots,d\}$.
The gaussian copula is a popular choice for the modeling of linear dependence structures.
The gaussian copula is constructed by extracting the dependence structure of the multivariate normal distribution and filtering the marginal influences,
\begin{equation}
C^{\text{gaussian}}[u_1, \hdots, u_d] = \Phi_{\Sigma}
[\phi^{-1}(u_1), \hdots, \phi^{-1}(u_d)],
\label{gaussian_copula}
\end{equation}
where $\phi$ is the cumulative distribution function of the standard normal distribution and $\Phi_{\Sigma}$ is the $d$-variate cumulative distribution function of the normal distribution with correlation matrix $\Sigma$. The correlation matrix $\Sigma \in [0,1]^{d \times d}$ contains $\frac{d(d-1)}{2}$ dependence parameters, $\rho_1, \hdots, \rho_{\frac{d(d-1)}{2}}$, governing the linear dependencies among the random variables $U_1, \hdots, U_d$. The density of a bivariate gaussian copula with dependence parameter $\rho = 0.4$ is displayed in the upper left panel of Fig.~\ref{copula_simulations}. The density only displays a linear relation between the variables. Similar to recovering linear dependence structures from the multivariate normal distribution, heavy-tailed dependence structures can be recovered from the multivariate students t-distribution using the t-copula
\begin{equation}
C^t[u_1, \hdots, u_d] = t_{\Sigma, \nu}[t_{\nu}^{-1}(u_1), \hdots, t_{\nu}^{-1}(u_d)].
\label{t_copula}
\end{equation}
where $t_\nu$ is the cumulative distribution function of the students t-distribution with degree of freedom $\nu$ and $t_{\Sigma, \nu}$ is the cumulative distribution function of the multivariate t-distribution with correlation matrix $\Sigma$ and degree of freedom $\nu$. Incorporating the degree of freedom $\nu \in (0, \infty)$ permits heavy tailed dependence structures. The heavy-tailedness can be interpreted as extreme events coinciding. A lower degree of freedom $\nu$ implies heavier tails. The density of a bivariate t-copula with dependence parameter $\rho = 0.4$ and degree of freedom $\nu = 4$ is displayed in the upper right panel of Fig.~\ref{copula_simulations}. The density displays the linear relation between the variables as well as the coincidence of extreme events. 
Another class of dependence structures can be described as asymmetric dependence structures. Two relevant copulas are the Gumbel and Clayton copula. The Gumbel copula exhibits dominant upper tail dependence while the Clayton copula exhibits dominant lower tail dependence. Their bivariate densities are displayed in the lower left, respectively lower right panel of Fig.~\ref{copula_simulations}. Both copulas are part of the archimedean copula family. Hence they are constructed as $C[u_1, \hdots, u_d] = \Psi^{-1}\left(\Psi(u_1) + \hdots + \Psi(u_d)\right)$ with a suitable generator function $\Psi$ \cite{genest1993statistical}, \cite{hofert2008sampling}. The generators for the Gumbel, respectively Clayton copula are given by
\begin{eqnarray}
\Psi^{\text{Clayton}}(t) =& (1 + t)^{- \frac{1}{\theta}},
\\
\Psi^{\text{Gumbel}}(t) =& e^{-t^{\frac{1}{\theta}}}.
\end{eqnarray}
The dominant lower tail dependence of the Clayton copula can be interpreted as lower tail events coinciding more often than upper tail events and vice versa for the dominant upper tail dependence of the Gumbel copula. An even more flexible copula model is the vine copula model. Vine copula models will be explained next.
\subsection{Vine Copulas}
Vine copulas are special pair copula constructions. The idea of pair copula constructions amounts to decomposing a $d$-variate dependence structure into a product of bivariate copulas. The joint density of $d$ random variables can, by virtue of the law of total probability, be decomposed into a product of conditional densities. Using the relation between conditional densities and copula densities (Eq.~\ref{conditional_density_copula}), one possible decomposition can be derived as \cite{aas2009pair, czado2010pair},
\begin{equation}
c[u_1, \hdots, u_d] = \prod_{j=1}^{d-1} \prod_{i=1}^{d-j} c[u_i, u_{i+j}|u_{i+1},\hdots,u_{j-1}].
\label{D_Vine}
\end{equation}
The decomposition is not unique. The decomposition in Eq.~\ref{D_Vine} is called drawable vine (D-vine). The unconditional copulas in the product all capture the dependence structure of neighboring variables, e.g. $c[u_i, u_{i+1}], c[u_{i+1}, u_{i+2}]$ and so forth. In a graphical representation, the connection between the variables resembles a straight line, hence the name D-vine.\footnote{Another special class of decompositions are canonical vines (C-vine). In this decomposition, the unconditional dependence structures are all centered around one variable, e.g. $c[u_i,u_{i+1}], c[u_i, u_{i+2}], \hdots$. In a graphical representation, the unconditional connection between variables resembles a star. In this work only D-vines are used.} The vine copula approach allows for flexible dependence modeling. It is advantageous in contexts where the bivariate dependence structures between variables can take different shapes, e.g. the dependence structure between variable $U_1$ and $U_2$ is linear whereas the dependence structure between $U_2$ and $U_3$ is heavy-tailed and so forth. Vine copula models can be estimated by maximum likelihood, we refer to \cite{aas2009pair} for details.
\subsection{Modeling Time Series with Spatio-Temporal Copulas}
\begin{figure}[!b]
\centering
\includegraphics[scale=0.5]{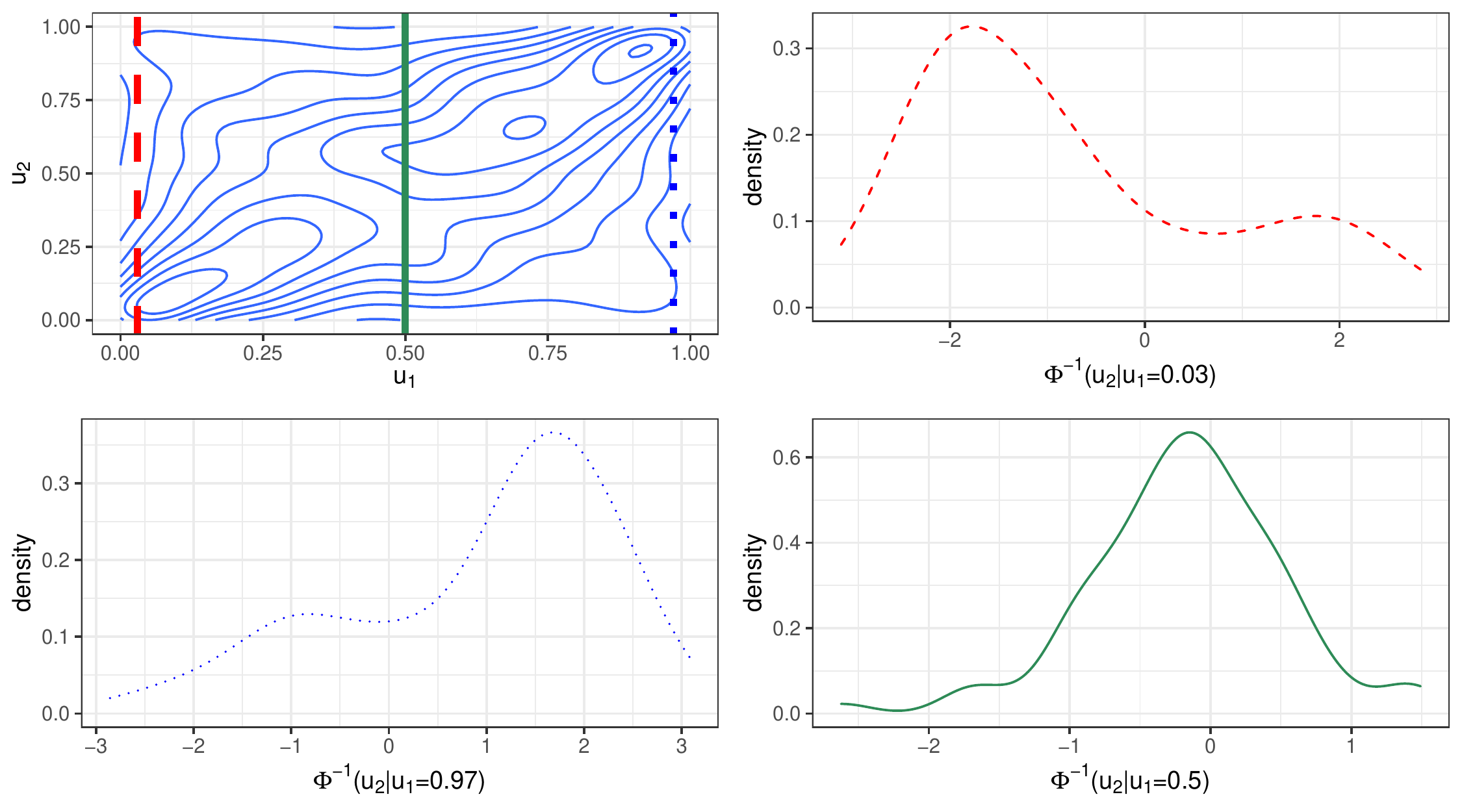}
\caption{Visualization of the conditional density structure depending on the value of the conditioning variable. The underlying model assumes the t-copula with dependence parameter $\rho = 0.4$ and degree of freedom $\nu = 2$. The marginal distribution is assumed as the standard normal distribution. The upper left panel shows the copula density of 2000 realizations of the before mentioned t-copula. The three lines indicate the three cases where the conditinal density is examined. The conditional density is calculated by aggregating all values in the neighborhood ($u_1 \pm 0.025$) of the conditioning variable and quantile-transforming them. The density in the upper right panel displays the conditional density given $u_1 = 0.03$. The lower panels display the conditional densities given $u_1 = 0.5$, respectively $u_1 = 0.97$.}
\label{Querschnitte}
\end{figure}

In this subsection, the copula-based time series models will be reviewed and summarized. It will be explained how these models can be used for forecasting.
First, the temporal copula modeling (see for example \cite{chen2006estimation}, \cite{beare2010copulas}) will be introduced. Eventually the combination of cross-sectional and temporal copula modeling, the spatio-temporal copula modeling, will be introduced. The exposition is based on the spatio-temporal t-copula modeling from \cite{simard2015forecasting} and vine copula modeling from \cite{beare2015vine, nagler2022stationary, smith2010modeling}. Further it will be examined how conditional temporal copula models offer a new approach to conditional heteroskedasticity modeling. The emergence of non-elliptical conditional distributions, respectively probabilistic forecasts, will be exemplified. The consequences for forecasting and the need for new point forecasting methods will be discussed.\\
Let $X_t$ be a univariate stationary Markov$(1)$ time series. The temporal evolution of the time series is completely specified by the joint distribution of random variables from consecutive time points i.e. $F_{X_t,X_{t-1}}$. Using Sklars theorem (Eq.~\ref{Sklar}), the joint distribution can be decomposed into copula and marginal distributions,
\begin{equation}
F_{X_{t}, X_{t-1}}(a, b) = C\left[F_{X_t}(a), F_{X_{t-1}}(b)\right].
\end{equation}
By the stationarity of $X_t$, $F_{X_t} = F_{X_{t-1}} =: F_X$. Hence the model can be determined by choosing an appropriate marginal distribution $F_X$ and an appropriate copula specification. Note that the marginal distribution $F_X$ is the unconditional distribution of $X_t$. Conditional properties of the time series are completely determined by the conditional copula. The conditional density of $X_{t}|X_{t-1}$ is given by 
\begin{equation}
f_{X_{t}|X_{t-1}}(a|b) = c\left[F_X(a), F_X(b)\right] f_{X}(a).
\end{equation}
Hence, for forecasting, the conditional density of $X_t|X_{t-1} = x_{t-1}$ can be used as probabilistic forecast.
This model can be understood as a generalization of the AR$(1)$ model \cite{smith2010modeling}\footnote{The generalization to $AR(p)$ models can be achieved by permiting the time series to be a Markov$(p)$ process.}. The gaussian autoregressive model can be recovered by choosing $C = C^{\text{gaussian}}$ and $F_X = \Phi$. When allowing other dependence structures, any temporal dependency representable by a copula can be reproduced. The concept of the copula based time series models can be further illustrated by its conditional model equation,
\begin{equation}
X_t|(X_{t-1}=x_{t-1}) = F_X^{-1}(C^{-1}\left[u_t | F_X(x_{t-1})\right]), \ \ \ u_t \sim U[0,1].
\end{equation}
In this formulation, the non-linear connection between $X_t$ and $X_{t-1}$ becomes obvious.
%
%
The generalization of the temporal copula time series model to $d$-variate time series, hence spatio-temporal time series models, is straight forward. Let $\mathbf{X}_t = (X_{1,t}, \hdots, X_{d,t})$ be stationary Markov$(1)$ time series. The structure of the time series is completely captured by the joint distribution of $\mathbf{X}_t$ and $\mathbf{X}_{t-1}$, 
\begin{equation}
F_{\mathbf{X}_t, \mathbf{X}_{t-1}}(\mathbf{a}, \mathbf{b}) = C\left[F_{X_{1}}(a_1),\hdots,F_{X_{d}}(a_d), F_{X_{1}}(b_1), \hdots, F_{X_{d}}(b_d)\right].
\end{equation}
The conditional density given observations from time point $t-1$ is as follows,
\begin{eqnarray}
f_{\mathbf{X}_t|\mathbf{X}_{t-1}}(\mathbf{a}|\mathbf{b}) =& \frac{c\left[F_{X_{1}}(a_1),\hdots,F_{X_{d}}(a_d), F_{X_{1}}(b_1), \hdots, F_{X_{d}}(b_d)\right]}{c\left[F_{X_{1}}(b_1), \hdots, F_{X_{d}}(b_d)\right]}
\\
& \times f_{X_1}(a_1) \cdot \hdots \cdot f_{X_d}(a_d). \nonumber
\end{eqnarray}
To sample from the conditional distribution, as necessary for Monte-Carlo approximations of conditional forecasts, the following procedure is employed \cite{simard2015forecasting}. First transform the observations at time $t-1$ to pseudo-observations. This is done by applying the probability integral transform to the observations, $(u_{t-1,1}, \hdots, u_{t-1,d}) := (F_{X_1}(x_{t-1,1}), \hdots, F_{X_d}(x_{t-1,d}))$. Then sample $n$ $d$-dimensional realizations from the conditional copula, Eq.~\ref{Conditional_Copula}. (Details on how to sample from the t-copula can be found in \cite{simard2015forecasting}. Details to sampling from vine copulas can be found in \cite{aas2009pair}). At last, the $n$ realizations have to be quantile transformed with their respective marginal distribution, yielding the $n$ samples of the conditional distribution. Relevant models for this work are the following. First, the spatio-temporal time series model where the copula is specified as the gaussian copula (Eq.~\ref{gaussian_copula}). The marginals are approximated non-parametrically by the empirical distribution.
\begin{eqnarray}
F_{\mathbf{X}_t, \mathbf{X}_{t-1}}(\mathbf{a}, \mathbf{b}) =& \Phi_{\Sigma}[\phi^{-1}(F^{\text{emp}}_{X_1}(a_1)), \hdots \phi^{-1}(F^{\text{emp}}_{X_d}(a_d)),
\\
& \phi^{-1}(F^{\text{emp}}_{X_1}(b_1)), \hdots, \phi^{-1}(F^{\text{emp}}_{X_d}(b_d))]. \nonumber
\label{S-Tem_gaussian}
\end{eqnarray}
This model is sensible to use when the dependence structure between the variables as well as the temporal dependence is linear. When the dependence strucure exhibits heavy-tailedness, the spatio-temporal t-copula model with non-parametric marginals poses a viable option,
\begin{eqnarray}
F_{\mathbf{X}_t, \mathbf{X}_{t-1}}(\mathbf{a}, \mathbf{b}) =& t_{\nu, \Sigma}[t_{\nu}^{-1}(F^{\text{emp}}_{X_1}(a_1)), \hdots t_{\nu}^{-1}(F^{\text{emp}}_{X_d}(a_d)),
\\
& t_{\nu}^{-1}(F^{\text{emp}}_{X_1}(b_1)), \hdots, t_{\nu}^{-1}(F^{\text{emp}}_{X_d}(b_d))]. \nonumber
\label{S-Tem_t}
\end{eqnarray}
For more flexible modeling, the spatio-temporal D-vine copula with non-parametric marginals can be utilized. For convenience, the model is presented in terms of its joint density and with variables $(\mathbf{a}, \mathbf{b}) =: \mathbf{p}$
\begin{eqnarray}
f_{\mathbf{X}_t, \mathbf{X}_{t-1}}(\mathbf{p}) =& \prod_{j=1}^{2d-1} \prod_{i=1}^{2d-j}
c[F_{X_i}(p_i), F_{X_{i+j}}(p_{i+j})|F_{X_{i+1}}(p_{i+1}),\hdots, F_{X_{j-1}}(p_{j-1})] \nonumber
\label{S-Tem_D-Vine}
\\
& \times f_{X_1}(p_1) \cdot \hdots \cdot f_{X_d}(p_d) f_{X_1}(p_{d+1}) \cdot \hdots \cdot f_{X_d}(p_{2d}).
\end{eqnarray}
This model is sensible to use when the dependence between variables differs in its structure or when the temporal dependence differs from the cross-sectional dependence.
As for solely temporal modeling, the temporal t-copula is employed,
\begin{equation}
F_{X_t, X_{t-1}}(a,b) = t_{\nu, \Sigma}[t_{\nu}^{-1}(F^{\text{emp}}_X(a)), t_{\nu}^{-1}(F^{\text{emp}}_X(b))].
\label{Tem_t}
\end{equation}
The heavy-tailed temporal dependence that this model exhibits is suitable for conditional heteroskedasticity modeling as will be discussed next.\\
%
%
The conditional distributions, respectively probabilistic forecasts from (spatio)-temporal copula time series models can be non-elliptical because of non-linear influences of the conditioning variable. In the following the behavior of the conditional distributions will be examined with regard to the temporal t-copula with standard normal marginal distribution\footnote{The choice of the standard normal distribution is just for convenience. The example would still be valid with other marginal distributions, e.g. students t-distribution.}.
The emergence of non-elliptical conditional densities from the heavy-tailed t-copula is visualized in Fig.~\ref{Querschnitte}. When the conditioning variable takes moderate values around $u_1 = 0.5$ the resulting conditional density is approximately elliptical. However, when the conditioning variable takes extreme values e.g. $u_1 = 0.03$ and $u_1 = 0.97$, the conditional density becomes bimodular. Thus, depending on the value of the conditioning variable, the resulting conditional density can have fundamentally different structures. This behavior offers a new approach to conditional heteroskedasticity modeling. Instead of widening the conditional density as in GARCH models, the density gets bimodular. This can be viewed as a sensible approach to volatility because the extreme behavior in volatile phases is mirrored in this model: When the time series takes a very low value at time point $t-1$ it can be expected that the value at time point $t$ will either be also very low or very high. The variance at time point $t$ is increased nevertheless, but the mechanism for the increased variance is a new one. \\
The temporal t-copula approach to conditional volatility, however, holds a problem. When the conditional density is non-elliptical it is not clear what constitutes a sensible point forecast. The expectation value may not be suitable in extreme cases where the conditional density is bimodular because the expectation value will take a value which is less probable than e.g. the modes. Taking the mode as point forecast could be a solution. Another possible solution to the problem of point forecasting is to augment the forecast by a artificial neural network (ANN). The ANN predicts which quantile of the conditional distribution is best (in terms of MSE) to use as point forecast. The inputs of the ANN are past values of the time series and the last optimal quantiles. The ANN architecture used in this work is the basic multi-layer perceptron (MLP) structure. We refer to \cite{higham2019deep} for an introduction to the topic. \\
\section{Results}
\label{Results}
\begin{figure}[!b]
\centering
\includegraphics[scale=0.5]{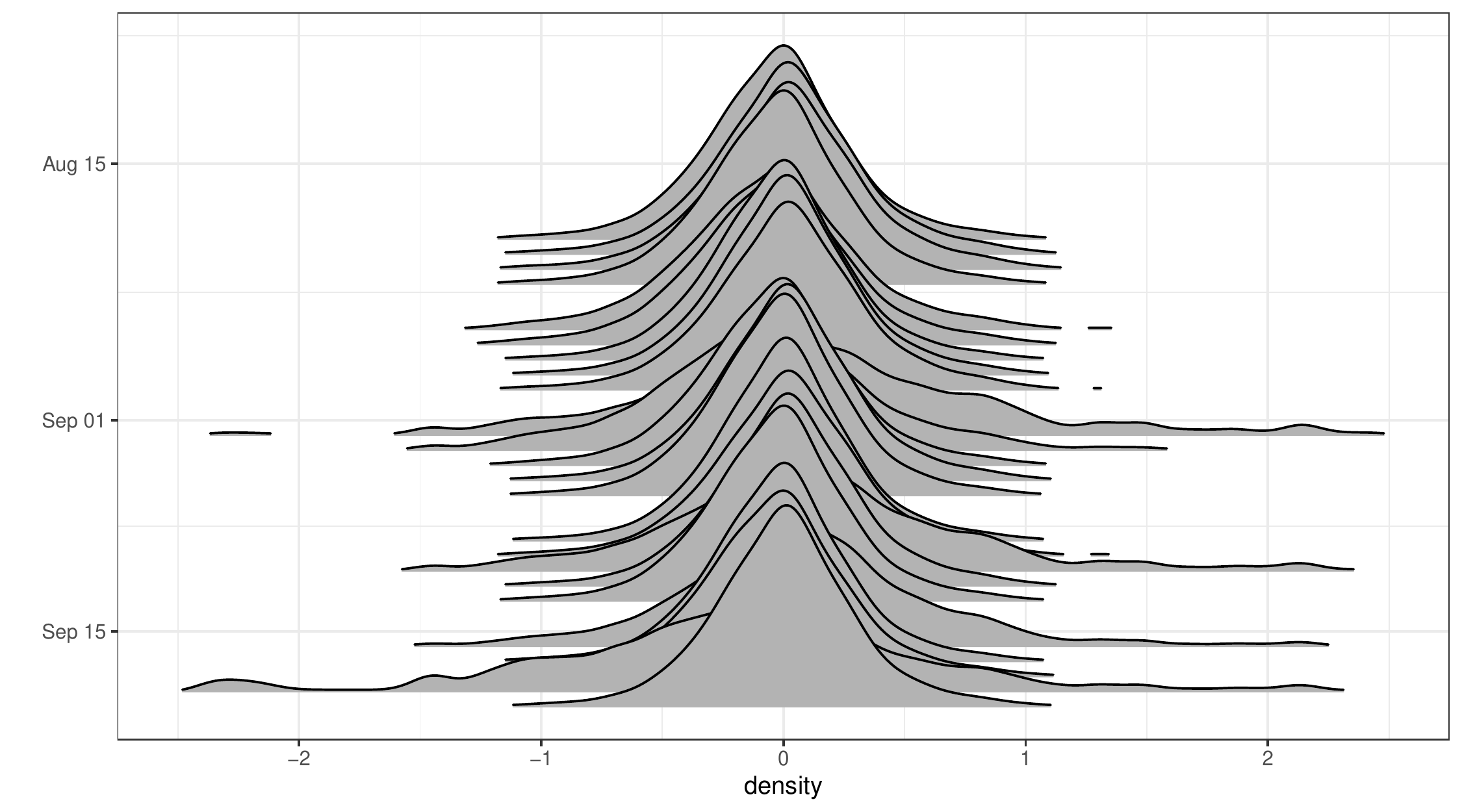}
\caption{Probabilistic forecasts for natural gas futures generated from the temporal t-copula model. The forecast densities can be non-elliptical during volatile times (August 2019 -- September 2019)}
\label{gg_shares}
\end{figure}

This section comprises the results of the expanding window forecasting study, investigating the performance of different models. The first 1000 observations (ranging from 2010-03-16
to 2013-12-08) are used as training data set. The following models are considered for evaluation.
\begin{itemize}
\item[1)] The temporal t-copula model with non-parametric marginals, Eq.~\ref{Tem_t}, henceforth denoted by Tem-t,
\item[2)] The spatio-temporal D-vine copula model Eq.~\ref{S-Tem_D-Vine}, henceforth denoted by S-Tem D-vine,
\item[3)] The spatio-temporal t-copula model, Eq.~\ref{S-Tem_t}, henceforth denoted by S-Tem-t,
\item[4)] The spatio-temporal gaussian copula model, Eq.~\ref{S-Tem_gaussian},  henceforth denoted by S-Tem-gaussian,
\item[5)] The Autoregressive moving average model with external regressors and absolute value, threshhold generalized autoregressive conditional heteroskedasticity model, henceforth denoted by ARMAX-AVTGARCH (closely related to the model from \cite{berrisch2022distributional}).
\end{itemize}
The models distributional forecasting performance is examined by the continous ranked probability score (CRPS) \cite{gneiting2007strictly}.
Further, the ANN assisted point forecasts of the S-Tem D-vine model and the Tem-t model are compared with the point forecasts from the ARMAX-AVTGARCH model. For each time series the ARMAX-AVTGARCH model is fitted individually.
All models are estimated via Maximum Likelihood. However, the marginals of the copula models are estimated non-parametrically to avoid transmitting estimation errors \cite{patton2013copula}. The order of the variables in the S-Tem D-vine copula model is fixed as
\begin{equation}
\text{CEF -- coal -- oil -- NGas -- NGas lag -- oil lag -- coal lag -- CEF lag}.
\end{equation}
This order is chosen to enable the lagged natural gas price to directly interact with the non-lagged natural gas price.
The gaussian, Gumbel, Clayton and t-copula are allowed as bivariate copulas in the D-vine decomposition (Eq.~\ref{D_Vine}). The probabilistic forecasts of all models are approximated by Monte-Carlo simulations with 1000 samples for each forecast.
Table \ref{CRPS} displays the models performances in terms of the CRPS. The ARMAX-AVTGARCH model performs best with regard to univariate distributional forecasting. However, the S-Tem D-vine model, the S-Tem-t and the Tem-t model are competitive. The performance of the copula models may be enhanced, when the marginal distributions are modeled parametrically. The empirical marginal distributions of the copula may not capture all marginal features of the time series. 
More versatile copula models could be used to enhance the forecast. The conditional dependence modeling may only be able to capture parts of the conditional effects.
\begin{table}[!t]
\caption{Aggregated CRPS values of the competing models for their one day-ahead probabilistic forecast for the four commodities. The CRPS is evaluated for the period 2013-12-19 -- 2021-02-23, comprising 1861 obervations.}
\label{CRPS}
\begin{tabular}{p{2cm}p{1.8cm}p{2cm}p{1.8cm}p{1.8cm}p{1.8cm}}
\hline\noalign{\smallskip}
Model/\newline Commodity & S-Tem \newline D-Vine & ARMAX-AVTGARCH & Tem-t & S-Tem-t & S-Tem- \newline gaussian \\
\hline
Natural Gas & $0.236$ & $0.227$ & $0.230$ & $0.234$ & $0.234$
\\
Oil & $0.564$ & $0.548$ & $0.551$ & $0.559$ & $0.558$
\\
Coal & $0.400$ & $0.389$ & $0.392$ & $0.398$ & $0.397$
\\
CEF & $0.236$ & $0.222$ & $0.229$ & $0.234$ & $0.234$
\\
\hline
\end{tabular}
\end{table}
The probabilistic forecasts from the Tem-t model during a volatile period is displayed in Fig.~\ref{gg_shares}. During volatile times the probabilistic forecasts are non-elliptical. During these times the ANN-augmented point forecasts can be valuable. The point forecasting performance of the models can be found in Table \ref{RMSE}. The evaluation starts at the 2001st observation, because the first 1000 probabilist forecasts are used to train the ANN. The hybrid, ANN-augmented S-Tem vine and the ANN-augmented Tem-t model generate the best point forecasts. The point forecasts of the ARMAX-AVTGARCH model are competitive though.
Note that the ANN model used for forecasting is build according to the basic multi-layer perceptron architecture. It is not perfectly suitable for catching sequential patterns. Using recurrent neural networks, especially long short-term memory architectures could enhance the performance even more and could be subject to future research. Also incorporating a measure for the structure of the probabilistic forecast could enhance the performance. However, this would requiere more advanced architectures. 
\begin{table}[!t]
\caption{Aggregated RMSE values of the competing point forecasting procedures for the four commodities. The RMSE is evaluated for the period 2017-10-23 -- 2021-02-23, comprising 861 observations.}
\label{RMSE}
\begin{tabular}{p{2cm}p{1.8cm}p{2cm}p{1.8cm}p{1.8cm}p{1.5cm}}
\hline\noalign{\smallskip}
Model/ \newline Commodity & S-Tem \newline D-Vine \newline ANN & ARMAX- \newline AVTGARCH & S-Tem \newline D-Vine \newline Mean & S-Tem \newline D-Vine \newline Mode & Tem-t \newline ANN \\
\hline
Gas & $0.594$ & $0.600$ & $0.599$ & $0.597$ & $0.589$
\\
oil & $1.222$ & $1.222$ & $1.236$ & $1.246$ & $1.220$
\\
coal & $1.000$ & $0.999$ & $1.009$ & $1.002$ & $0.997$
\\
CEF & $0.605$ & $0.608$ & $0.614$ & $0.604$ & $0.607$
\\
\noalign{\smallskip}\hline\noalign{\smallskip}
\end{tabular}
\end{table}

\section{Conclusion}
\label{Conclusion}
The application of copula-based time series models to natural gas and related commoditiy prices is explored in this work. An expanding window forecasting study is conducted. The time series used for analysis are extracted from \url{investing.com} via the \texttt{Python} package \texttt{investpy}. The time series comprises short term future price series of natural gas, crude oil, coal and carbon emissions. \\
After introducing the basic notions of dependence modeling with copulas and the D-Vine copula, the copula based time series models from the literature are reviewed. The emergence of non-elliptical probabilistic forecasts is exemplified using the temporal t-copula. It is visualized how the temporal t-copula offers a new approach to conditional heteroskedasticity modeling. It is not clear what constitutes a sensible point forecast when the probabilistic forecast is non-elliptic. To this end a artificial neural network is employed to predict what quantile of the probabilistic forecast is best to use as point forecast. The inputs of the artificial neural network are past values of the multivariate time series and the last best quantiles of the probabilistic forecast. \\
In the forecasting study, the predictive performance of the temporal t-copula, the spatio-temporal t-copula and the spatio-temporal D-Vine copula is examined. The marginal distributions are estimated by the respective empirical distribution. The performance is compared with the performance of an autoregressive moving-average model with external regressors and absolute value, threshhold generalized autoregressive conditional heteroskedasticity modeling (ARMAX-AVTGARCH). A closely related model was recently shown to be the best model for natural gas forecasting. Hence it is understood as benchmark model. The distributional predicitive performance is examined by the continious ranked probability score (CRPS).
We find that the copula-based time series models are competitive with the ARMAX-AVTGARCH model. The point forecasts are evaluated by the root mean squared error (RMSE). The ANN-augmented point forecasts perform best, although the forecasts from the ARMAX-AVTGARCH model are still competitive. \\
The performance of the copula-based time series models could be enhanced by modeling the marginal distributions parametrically. The non-parametric modeling may not catch all  marginal features of the time series. However, this procedure requieres the estimation to be conducted in one step to guarantee efficient estimation. Another possibility to enhance the performance is to consider more versatile copula models. The current modeling may not capture all conditional features of the time series. Another possibility, with regard to the vine copula model, is to consider other vine structures. In this work the D-vine structure was imposed. Other structures may be able to capture the dependencies better.
As for the point forecasts, it was shown that the ANN-augmented forecasts perform well. Even though we choose to utilize the standard multi-layer perceptron architecture, which can not model sequential information perfectly well, the precision was increased. Using more sophisticated architectures that are more suited to catch sequential information will be subject to future research. It would also be interesting to use other models to predict the best quantile for point forecasting.

\vspace*{1cm}
\textbf{Acknowledgement:}\\
The authors gratefully acknowledge the computing time
provided on the Linux HPC cluster at Technical University
Dortmund (LiDO3), partially funded in the course of the
Large-Scale Equipment Initiative by the German Research
Foundation (DFG) as project 271512359.

\input{bibi}

\end{document}

%% file: bibi.tex
%
%
%